\documentclass[a4paper,english,preprint,showpacs]{revtex4}

\usepackage{pslatex}
\usepackage{babel}

\usepackage{amsmath}
\usepackage{amssymb,verbatim}

\usepackage{array}
\usepackage{longtable}
\usepackage{subfigure}
\usepackage{graphicx}

\begin{document}
\title{The transition from highly to fully stretched polymer brushes in good solvent}

\author{Ivan Coluzza}

\affiliation{Department of Chemistry, University of Cambridge, Lensfield Road, Cambridge CB2 1EW, UK.}

\author{Jean-Pierre Hansen}
\affiliation{Department of Chemistry, University of Cambridge, Lensfield Road, Cambridge CB2 1EW, UK.}


\date{\today}

\begin{abstract}
The stretching of brushes of long polymers grafted to a
planar surface is investigated by Monte Carlo 
simulations in the limit of very high grafting densities, as achieved in recent experiments.
The monomer density profiles are shown to deviate considerably from the 
parabolic limiting form predicted by self-consistent field theory.
A rapid transition is observed from parabolic to fully stretched
polymers, characterized by a dramatic change in the end-monomer height distribution and by a clear
cross-over in the slope of the brush height versus scaled grafting density.

\end{abstract}

\maketitle
There have been numerous experimental~\citep{Auroy91}, theoretical~\citep{Gennes80,Milner88,Netz98} and simulation~\citep{Binder02,Kreer04} studies of polymer
``brushes'' grafted to substrates, confirming the Alexander
scaling of the brush height with grafting density~\citep{Alexander77}. They show that the monomer density
profiles go over to a universal parabolic profile as a function of the distance from the
surface, for sufficiently strong stretching, as predicted by self-consistent field theory (SCF)~\citep{Milner88,Netz98,Seidel00}. Recent experimental techniques based on surface-initiated polymerization~\citep{Jones02,Ohno05} allow higher grafting densities than previously achieved. In this letter we analyse the break-down of SCF theory for long chains and high grafting
densities by advanced Monte Carlo simulations of polymers in good solvent. Upon increasing the grafting
density, we observe a transition from the parabolic regime to fully stretched
brushes which correlates with a dramatic change in the end-monomer density profile. The
exposure of the end monomers of very dense brushes opens new possibilities
for the development of chemically active soft surfaces.

Consider a layer of N identical
flexible polymers of $M$ monomers and contour length $L=(M-1)b$ (where $b$ is the segment length), in good solvent. The
first monomer of each chain is chemically grafted to a grafting site on a planar substrate of area A; the dimensionless
grafting density is defined as $\sigma=Nb^2/A$. For convenience b will henceforth be the unit length. The key length
scales are the mean distance between grafting sites, $d\sim \sigma^{-1/2}$, the radius of gyration of an isolated polymer,
 $R_{g}\sim L^\nu$ (where the Flory exponent $\nu = 0.588$ under good solvent conditions) and the brush height h which, according to the
Alexander-de Gennes mean field argument~\citep{Gennes80,Alexander77}, scales like 
\begin{equation}
h\sim L \sigma^{\frac{1-\nu}{2\nu}}\sim L \sigma^{0.35} \sim L d^{-0.7} \label{Heigth}
\end{equation}
i.e. the chains stretch upon increasing the grafting density due to their mutual repulsion. This scaling
has been confirmed by neutron scattering diffraction experiments~\citep{Auroy91}. An important dimensionless
parameter is $\sigma_{g}=\sigma R_{g}^{2}\sim R_{g}^2/d^2$. When $\sigma_{g}<1$, polymers grafted to neighbouring sites do not overlap and the
height $h$ of quasi-independent coils scales like $h\sim R_{g}\sim L^\nu$ (``mushroom regime''). When $\sigma_{g}>1$, neighbouring coils interact and begin
to stretch, i.e. $h\sim L$. This regime has been widely studied within SCF theory~\citep{Gennes80,Milner88,Seidel00} and by MC simulations of various
models, involving in general rather short chains ($L<100$)~\citep{Murat89,Lai91,Binder02,Kreer04,Seidel00}. The polymer ``brush'' may be characterized by the
monomer distribution function $P(z)$, where $z$ is the vertical distance from the substrate, normalized such that
$\int_{0}^L P(z)dz=N$. Within SCF, the dimensionless control parameter is $\beta=(h/R_{go})^2$ (where $R_{go}\sim L^{1/2}$ is
the radius of gyration of the non-interacting, Gaussian chain), and in the strong stretching regime $\beta>>1$, SCF
reduces to its ``classical'' limit, which predicts a universal parabolic curve when $P(z)\sigma^{(1-\nu)/\nu}$ is plotted
versus $z/\left(L \sigma^{(1-\nu)/2\nu})\right)$~\citep{Milner88,Netz98}. 
\begin{figure}
\begin{center}
\includegraphics[width=1.00\columnwidth]{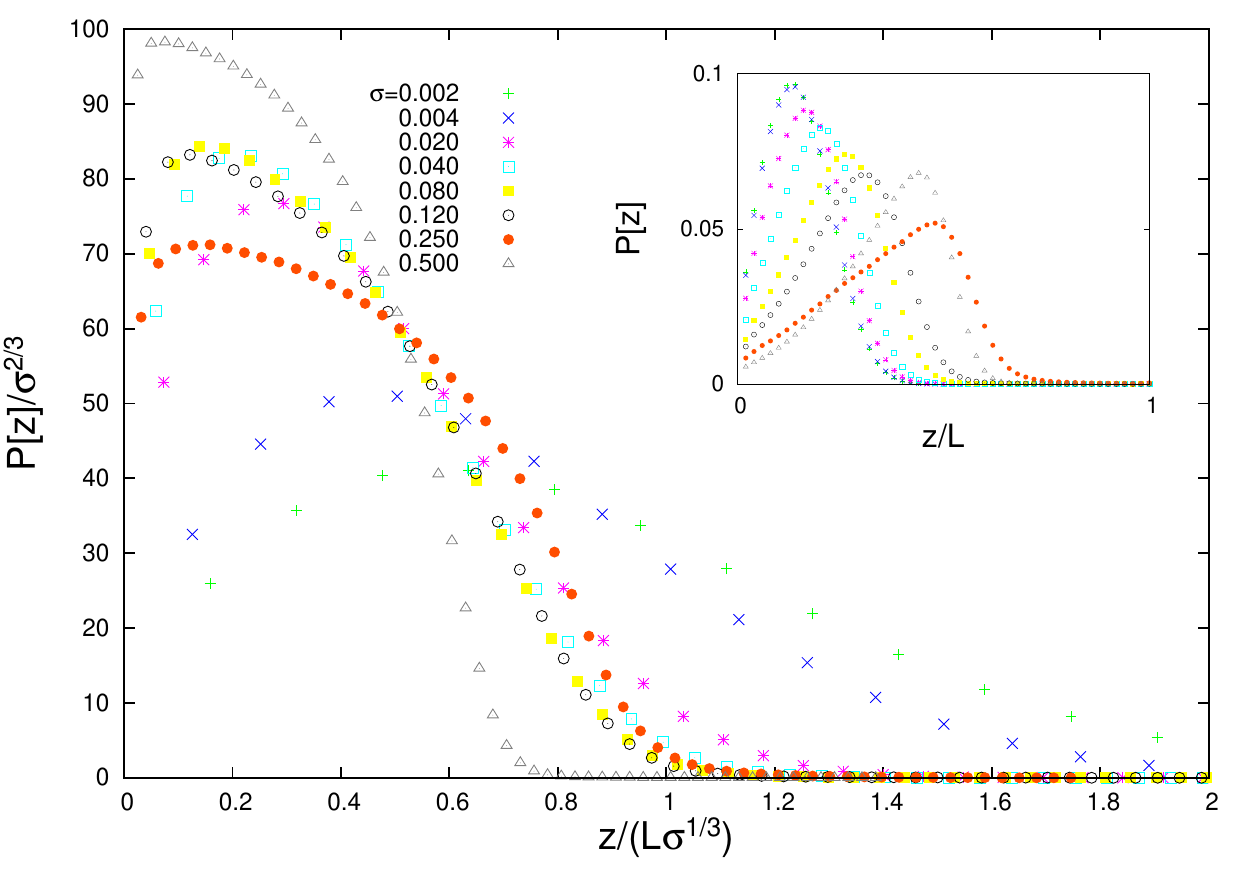}
\end{center}
\caption{(Color on-line)Scaled monomer distribution function versus scaled altitude for brushes of SAW
polymers of length $L=50$, and several grafting densities $0.002\leq \sigma \leq 0.5$. The two lowest $\sigma$ 
correspond to the ``mushroom'' regime, while deviations from SCF scaling occur for $\sigma \geq 0.25$.
The inset shows the end monomer altitude profiles versus $z/L$ for the same values of $\sigma$ \label{Fig_1}}

\end{figure}
This scaling is confirmed by MC simulations, with deviations from the parabolic shape at small and large $z$ in the case of
moderate stretching~\citep{Binder02,Kreer04,Seidel00,Murat89,Lai91}. Some deviations from a universal master curve, due to a flattening of the profiles
at intermediate heights, have been observed in simulations at the highest grafting densities $\sigma_g$ explored so far
~\citep{Seidel00,Pal06,HeMerlitzMacromolecules07}. 
 Meanwhile experimental grafting techniques based on surface-initiated polymerization~\citep{Jones02} have achieved
extremely high grafting densities, up to nearly one site per $nm^2$~\citep{Jones02,Ohno05}.

To explore the structure of polymer brushes
in the regime of such high grafting densities, corresponding to the regime of $\sigma_g>>1$, which may be achieved by
increasing the grafting density $\sigma$ or considering very long polymers, we have carried out extensive MC simulations
of the simplest lattice model of flexible polymers in good solvent, namely self (and mutually) avoiding walks (SAW) on
a cubic lattice. The grafting sites are on a square lattice of identical spacing $b$. We used ``annealed'' grafting
conditions whereby the first monomer of each chain can jump between nearest neighbor grafting sites in order to speed
up equilibration~\citep{Lai91}. Three types of MC moves
were used: discrete translations of entire chains on the square lattice in the $x-y$
grafting plane; chain re-growth moves based on the Configurational Bias Monte Carlo
(CBMC) algorithm~\cite{FrenkelSmit} ; segment re-growth moves based on CBMC with fixed
end-points~\cite{FrenkelSmit}. This combination of moves is expected to 
guarantee good equilibration of the simulated brush up to very high grafting densities
(corresponding to moderately high monomer volume fractions in the brush).
The initial conditions were generated with fully stretched polymer conformations
attached to randomly chosen grafting centres on the square lattice.
The grafting centres were allowed to diffuse on the square lattice (annealed sampling),
which is equivalent to a random sampling of fixed grafting points, followed by a 
statistical average over the randomly chosen initial configuration~\cite{Lai91}.
After initial equilibration, statistical averages of the monomer and end-monomer profiles were taken over
typically $2\cdot 10^8$ MC moves. 

\begin{figure}
\begin{center}
\includegraphics[width=1.00\columnwidth]{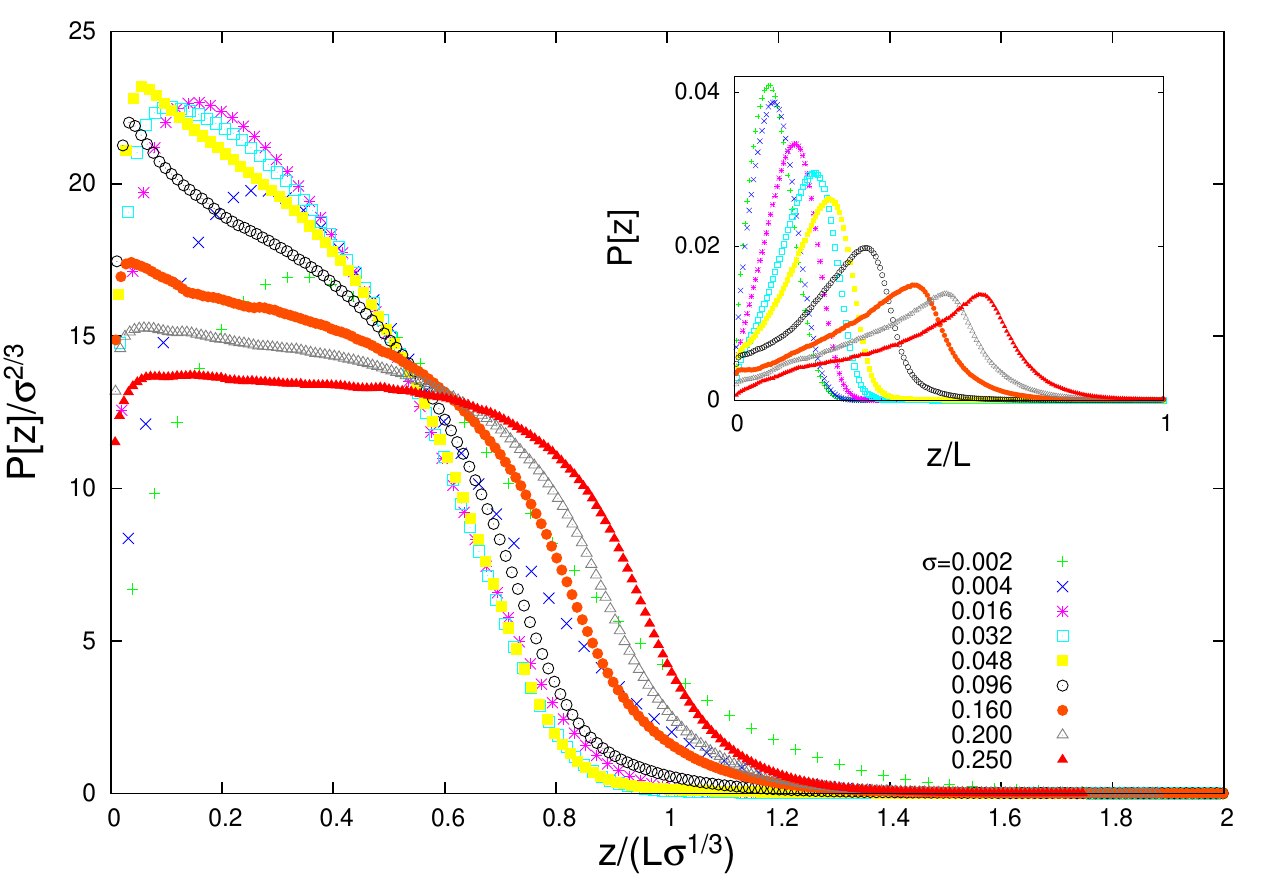}
\end{center}
\caption{(Color on-line)Same as in Fig.~\ref{Fig_1}, but for $L=200$ and  $0.002\leq \sigma \leq 0.25$.
Increasing deviations from SCF scaling regime are observed for $\sigma > 0.05$\label{Fig_2}}
\end{figure}

\begin{figure}
\begin{center}
\includegraphics[width=1.00\columnwidth]{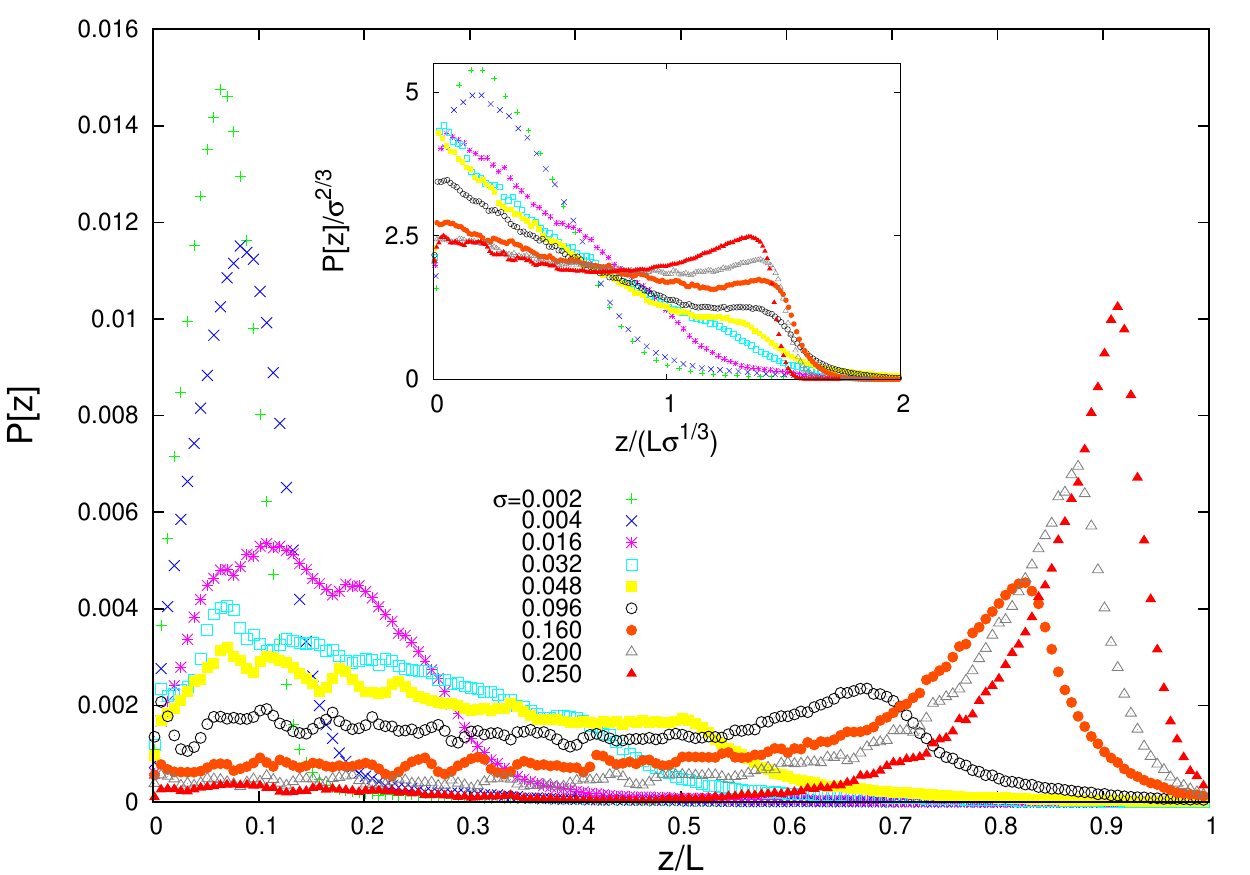}
\end{center}
\caption{(Color on-line)End-monomer height distribution function of grafted SAW polymers of length $L=800$ versus $z/L$ for grafting densities $0.002 \leq \sigma \leq 0.25$.
Note the rather sharp transition to full stretching when $\sigma \simeq 0.05$, which contrasts with the gradual stretching observed 
for shorter chains in Fig.~\ref{Fig_1}-\ref{Fig_2}. Inset: scaled monomer distribution function versus scaled altitude  for the same
values of $\sigma$. Deviations from the SCF scaling regime set in for $\sigma \gtrsim 0.01$, and the profiles become nearly rectangular
(full stretching) at the highest grafting densities.\label{Fig_3}}
\end{figure}

We have systematically
computed the monomer and end-monomer distribution functions $P(z)$ and $P_{M}(z)$ over a wide range of $\sigma$, and for lengths
$L=$50, 200, 400 and 800. Scaled profiles for short chains ($L=50$) and several $\sigma$ are shown in Fig.~\ref{Fig_1}. The universal
 scaling regime is reached for $\sigma>0.04$ ($\sigma_g>0.6$), and persists up to $\sigma=0.25$ ($\sigma_g\simeq 4$). Below $\sigma\simeq 0.04$
($\sigma_{g}<0.6$), interactions between neighbouring grafted polymers become negligible so that the ``mushroom'' regime is
reached, and the profiles no longer follow the SCF scaling. The end-monomer profile $P_{M}(z)$ moves to higher $z$ as
$\sigma$ increases, as one might expect. The probability of the end monomer ``returning'' towards the grafting surface
decreases with increasing $\sigma$~\citep{Seidel00}. Fig.~\ref{Fig_2} shows similar results, but for longer chains ($L=$200). The mushroom regime
is only observed  for $\sigma<0.004$ ($\sigma_{g} <0.3$), while the scaling regime is satisfied for $0.01\leq \sigma \leq 0.05$
($0.8 \leq \sigma_{g} \leq 4$). For $\sigma > 0.05$, the profiles deviate increasingly from the scaling prediction, and for $\sigma > 0.2
$($\sigma_{g} > 16$) $P(z)$ flattens and stretches considerably. Concomitantly the end-monomer profiles $P_{M}(z)$ move further and
further towards their upper limit, $z=L$. These trends are considerably enhanced in the case $L=800$, as shown in Fig.~\ref{Fig_3}.
The SCF scaling regime only holds for $\sigma<0.02$ ($\sigma_{g}<10$) while for higher $\sigma$, $P(z)$ flattens and stretches
towards the rectangular profile ($P(z)=1$, $z<L$; $P(z)=0$, $z>L$) which is the exact limit for the lattice model, when $\sigma=1$
(each polymer then reduces to a rod of length $L$ because of the excluded volume constraint). This strong stretching
behavior correlates well with the behavior of $P_{M}(z)$, which undergoes a dramatic transition around $\sigma=0.1$
($\sigma_{g}=40$), when the peak in the end-monomer distribution function sharpens and moves towards $z=L$ (in the limit
$\sigma \rightarrow 1$, $P_{M}(z) \rightarrow \delta(z-L)$).

\begin{figure}
\begin{center}
\includegraphics[width=1.00\columnwidth]{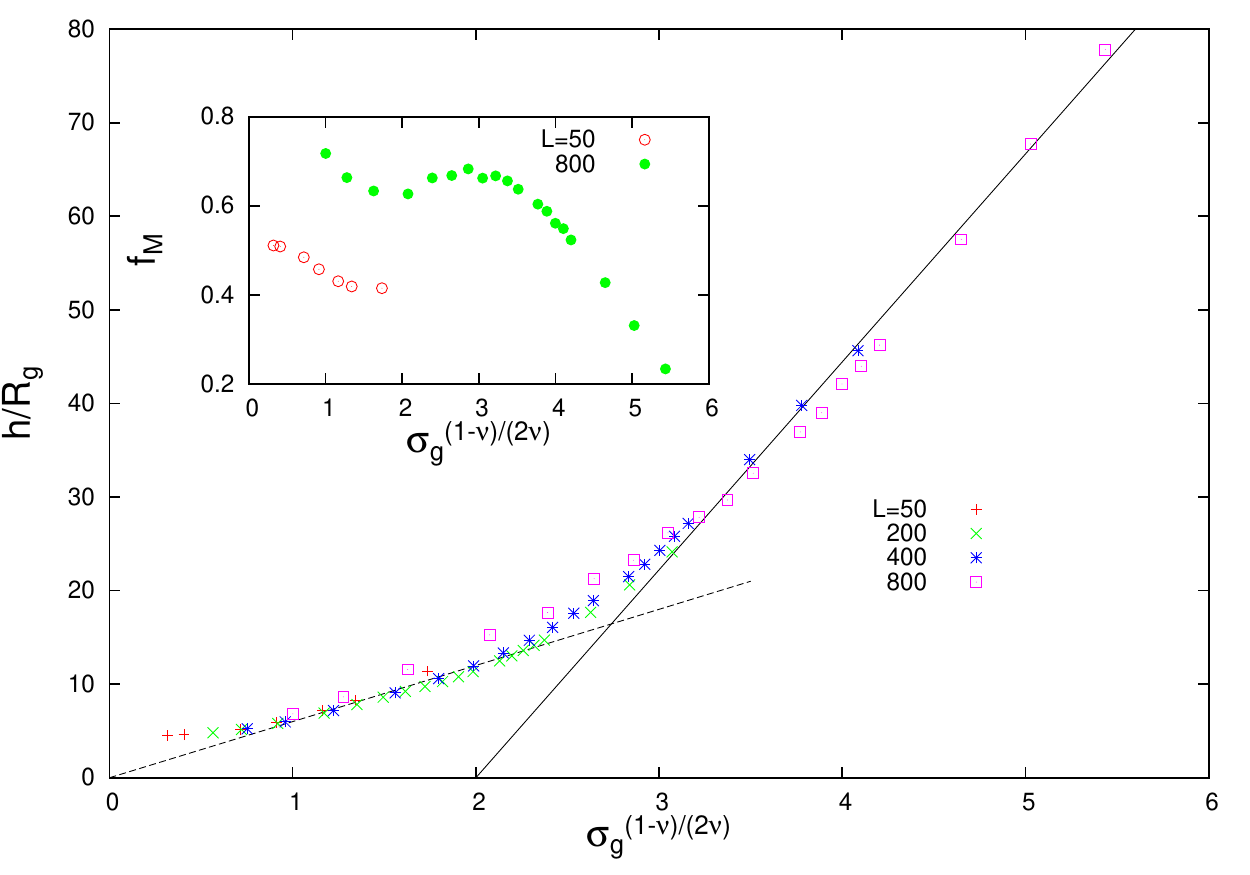}
\end{center}
\caption{(Color on-line)Plot of the mean height $h=<z_M>$ of a brush of SAW polymers (divided by $R_g$), versus $\sigma_g^{(1-\nu)/2\nu}$ for
$L=50$,$200$,$400$ and $800$. The slopes of the two straight lines fitted to the low $\sigma_g$ and high $\sigma_g$ data
indicate a cross-over from the SCF regime to the fully stretched regime around $\sigma_g \simeq 10$. The slope
($\simeq 5.2$) of the low $\sigma_g$ regime (dashed line) is close to earlier estimates obtained for different 
polymer models~\cite{Binder02,Kreer04}. Inset: relative fluctuation $f_M=\left[ \left<z_{M}^2\right> - \left<z_{M}\right>^2\right]^{1/2}/\left<z_{M}\right>$ 
of the brush height versus $\sigma_g^{(1-\nu)/2\nu}$ for $L=50$ and $L=800$, The fluctuations
drops rapidly beyond the cross-over density, pointing to the homogeneity of the highly stretched brush.  \label{Fig_4}}
 \end{figure}

The mean height $h$ of the brush is the first moment of the probability density $P(z)$; the MC results for the ratio $h/R_g$
are plotted in Fig.~\ref{Fig_4} vs $\sigma_g^{(1-\nu)/2\nu}$ for $L=200$, $400$ and $800$. The data show a clear cross-over between two linear regimes with different slopes around
 $\sigma_g\simeq 10$. The linearity agrees with the prediction of Alexander-de Gennes scaling~\citep{Auroy91}, but the change in slope suggests a cross-over
 from the SCF regime to the regime dominated by strong
 excluded volume correlations which enhance stretching of the chains.
 Although the slope ($\simeq 5.2$) of the low $\sigma_g$ regime is close to that of the earlier 
 estimates obtained for different polymer models~\citep{Binder02,Kreer04}, a clear-cut cross-over to the strong
 stretching regime has not been reported earlier. Indications in that direction are contained in the early work of 
Grest et~al.~\citep{Grest94,Grest93}. By using a continuous model of polymer brushes, they observed deviations from the SCF scaling regime
for grafting densities of 0.15 and chains 200 monomers long, resulting in enhanced stretching of the brush. However there
was no clear evidence of a sharp transition between the SCF and the fully stretched scaling regimes, similar
to to that shown in the end-monomer distribution function (Fig.~\ref{Fig_3}), or in the scaling plot (Fig.~\ref{Fig_4}) of the mean brush height. This was probably due to insufficient polymer length and to the presence of an attractive potential between the monomers.
 
Let us know consider the ``roughness'' of the brush/solvent interface may be characterized by the relative fluctuation $f_M$ of the
 end-monomer altitude, $\left[ \left<z_{M}^2\right> - \left<z_{M}\right>^2\right]^{1/2}/\left<z_{M}\right>$ which is readily derived from 
 $P_{M}(z)$. The MC results for $f_M$ are plotted in Fig.~\ref{Fig_4} versus $\sigma$ for L=50 and 800. The difference is dramatic:
 while $f_M$ saturates rapidly at a value close to 0.4 for the shortest chains,
 the fluctuation is seen to decrease steadily for the longer chains, pointing to a relatively smooth
 upper interface of the brush.
 It is worthwhile to note that the corresponding fluctuation of the center of mass (CM) of the chains follows very similar
 curves, slightly below those for $f_M$.

 Sampling of stretched polymer conformations at high grafting densities becomes increasingly difficult for long polymers,
 such that $\sigma_g> 10^2$. A way out is to switch to a multi-blob representation of the grafted chains,
 as recently proposed for homopolymers and block copolymers in the bulk~\citep{Pierleoni06,Capone07}. 
 Within this coarse-graining procedure each chain is divided into $n$ blobs of length
 $l=L/n$ and radius of gyration $r_g \sim l^\nu$. The average blob density within the volume
 of the brush of height $h$ is $\rho_b=N n/(A h) \sim N L h^{(1-3\nu)/2\nu}$, while the
 overlap density of blobs is $\rho_b^* \sim 1/r_g^3 \sim 1/l^{3\nu}$ so that 
 $\rho_b/\rho_b^* \sim (L/n\sigma^{1/2\nu})^{3\nu-1}$. The minimum number of blobs 
 required to ensure  that blobs do not, on average, overlap is hence
 $n \sim \sigma_g^{1/2\nu}$. Effective interactions between 
 the CM's of the bonded and non-bonded blobs and between a blob and the
 substrate, are determined by averaging over monomer degrees of freedom
 for given CM-CM distances.
 The coarse-grained multi-blob model requires effective pair potentials
$v(r)$ and $\phi_1(r)$ between the CM's of non-bonded and bonded
blobs on the same or different (in the case of $v(r)$) chains,
as well as an effective wall-blob potential $\psi(z)$, and an effective tethering
potential $\phi_2(r)$ between the CM of the first blob and the grafting centre.
All these effective interactions are determined by inverting MC results for the pair distribution functions between
the CM's of a single grafted polymer made up of a small number of blobs~\cite{Pierleoni06,Capone07,Bolhuis01},
and are assumed to be transferable to finite grafting density conditions, as long as the system is 
in the weak overlap ($\rho_b<\rho_b^*$) regime.
  \begin{figure}
\begin{center}
\includegraphics[width=1.00\columnwidth]{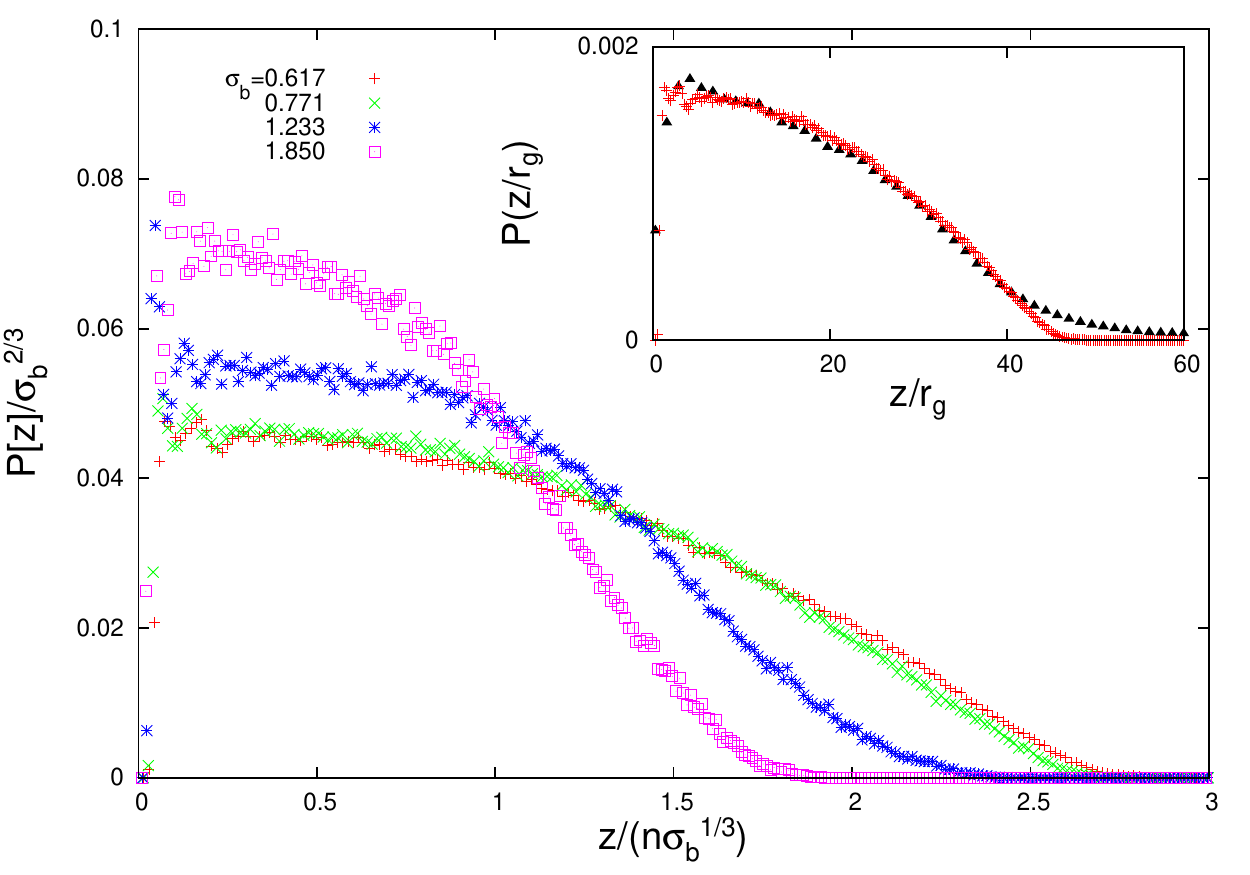}
\end{center}
\caption{(Color on-line)Scaled monomer distribution function versus scaled altitude for brushes of multi-blob polymers of length $n=20$
blobs, and  $1.8\geq \sigma_b \geq 0.6$ where $\sigma_b=\sigma\,r_g^2$. Increasing deviations from SCF scaling regime are observed for $\sigma_b>1.2$.
Inset: comparison between the scaled monomer distribution function of a SAW polymer of length $L=800$ at a grafting
density $\sigma=0.008$ (triangles)  and a multi-blob polymer under the corresponding physical condition of $\sigma_b=0.617$ (continuous curve). The plot
shows very good agreement between the two representations.\label{Fig_5}}
 \end{figure}
The form of the repulsive potential $v(r)$ is Gaussian to a good approximation 
\begin{equation}
\frac{v(r)}{k_B T}\simeq A e^{-\alpha \left(r/r_g\right)^2} \label {M1}
\end{equation}
with $A \simeq 1.8$ and $\alpha \simeq 0.8$. The tethering
potential $\phi_1(r)$ is well represented by the sum of $v(r)$
and a harmonic spring potential, while $\phi_2(r)$ is a similar
harmonic spring potential, and the wall-blob potential $\psi(z)$
is an exponential repulsion of range $\simeq r_g$~\cite{Bolhuis01}.
MC simulations of the multi-blob representation of grafted polymers sample
single blob and total chain displacements with a standard Metropolis acceptance
criterion, which are sufficient because of the softness of the above effective potentials.
 If $n$ is chosen such as to satisfy the weak overlap condition above,
 the effective interactions can be extracted from low density MC
 simulations. This coarse-graining leads to an
 enhancement of sampling efficiency by several orders of magnitude,
 because of the reduction of the total number of degrees of freedom by a factor $l$, and the softness of the effective interactions.
 MC results for the CM probability densities, as determined for coarse-grained polymer chains of
 $n=5$ and $20$ blobs are shown in Fig.~\ref{Fig_5}. The profiles are seen to obey the SCF scaling at intermediate
 grafting densities, and to exhibit the enhanced stretching at very high grafting densities, in good agreement with the full monomer-level 
 MC results of Fig.~\ref{Fig_1}-\ref{Fig_4}.

 In summary, we have shown by MC simulations of long grafted polymer chains that the brush
 profile and height undergo a rapid transition when the grafting
 density $\sigma_g=\sigma R_{g}^2 > 10$. Beyond that value, the probability 
 density $P(z)$ switches rapidly from universal,
 quasi-parabolic profile to a quasi-rectangular shape, while the fluctuations
 of the end-monomer
 altitude drops sharply, pointing to a rather flat brush/solvent interface. These changes are induced by correlations between monomers on neighbouring chains,
 which are neglected in SCF theory. We conjecture that these
 correlations will lead to an effective, entropic repulsion between stretched polymers,
 similar to the Helfrich interaction~\citep{Helfrich90} between stacked membranes, which in turn may induce an ordering of the 
 grafting centres on the substrate if these centres are mobile, as for copolymers anchored at a 
 liquid-liquid interface, just as the Helfrich repulsion leads to a lamellar phase of stacked membranes.
 Our results point to new possibilities in experimental realization of chemically active surfaces, because the height distribution
 of the end monomer controls the chemical activity of the brush. Moreover the activity of
 such soft surfaces could be controlled by the length of the grafted polymers.

\begin{acknowledgments} 
The authors are grateful to Barbara Capone for the her help
 with the calculation of effective blob-blob interactions, and to Wilhelm Huck for an
 enlighting discussion. IC acknowledges a
grant from the Marie-Curie FP6 program.
\end{acknowledgments}

\end{document}